\documentclass[twocolumn,tighten,times]{aastex6}
\usepackage{physics}
\usepackage{bm}
\usepackage[normalem]{ulem}

\shorttitle{Internal gravity waves. Energy Transport}
\shortauthors{Vigeesh et al.}

\begin{document}

\title{Internal Gravity Waves in the Magnetized Solar Atmosphere.\\
II.  Energy Transport}

\author{{G.~Vigeesh}\altaffilmark{1}, {M.~Roth}\altaffilmark{1}, 
{O.~Steiner}\altaffilmark{1,2}, and {J.~Jackiewicz}\altaffilmark{3}}
\affil{\altaffilmark{1}Leibniz-Institut f\"{u}r Sonnenphysik (KIS), 
Sch\"{o}neckstrasse 6, 79104 Freiburg, Germany\\
\altaffilmark{2}Istituto Ricerche Solari Locarno (IRSOL), via Patocchi
57-Prato Pernice, 6605 Locarno-Monti, Switzerland\\
\altaffilmark{3}New Mexico State University, Department of Astronomy, 
	P.O. Box 30001, MSC 4500, Las Cruces, NM 88003, USA
}

\email{vigeesh@leibniz-kis.de} 

\begin{abstract}
In this second paper of the series on internal gravity waves (IGWs), 
we present a study of the generation and propagation of IGWs in a 
model solar atmosphere with diverse magnetic conditions. A magnetic 
field free, and three magnetic models that start with an initial, 
vertical, homogeneous field of 10 G, 50 G, and 100 G magnetic flux 
density, are simulated using the {CO$^{\rm 5}$BOLD} code. We find that 
the IGWs are generated in similar manner in all four models in spite 
of the differences in the magnetic environment.  The mechanical energy 
carried by IGWs is significantly larger than that of the acoustic waves 
in the lower part of the atmosphere, making them an important 
component of the total wave energy budget. The mechanical energy
flux (10$^{6}$\textendash10$^{3}$ W m$^{-2}$) is few orders of 
magnitude larger than the Poynting flux 
(10$^{3}$\textendash10$^{1}$ W m$^{-2}$). The Poynting fluxes show a 
downward component in the frequency range corresponding to the IGWs, 
which confirm that these waves do not propagate upwards in the 
atmosphere when the fields are predominantly vertical and strong.  
We conclude that, in the upper photosphere, the propagation 
properties of IGWs depend on the average magnetic field strength and 
therefore these waves can be potential candidate for magnetic field 
diagnostics of these layers. However, their subsequent coupling to 
Alfv\'{e}nic waves are unlikely in a magnetic environment permeated 
with predominantly vertical fields and therefore they may not 
directly or indirectly contribute to the heating of layers above 
plasma-$\beta$ less than 1.

\end{abstract}


\section{Introduction}{\label{s:introduction}}

Atmospheric waves are part and parcel of any gravitating body that has
a gaseous envelope. Whether it be terrestrial or other planetary
atmospheres, or the solar atmosphere---a plethora of waves are found
in them that are responsible for energy and momentum transport across
its different layers. 
In the case of the Sun, these waves play an 
	important role in the overall dynamics of the solar atmosphere and 
	help to connect its lower layer---a region where a majority of the 
	waves supposedly originate, with its upper layers.

Internal gravity waves (IGWs) are archetype of atmospheric waves with
buoyancy acting as their main driving mechanism
\citep{2001wafl.book.....L,sutherland2010}. 
When compressibility
effects are also taken into account, a combination of IGWs and sound
waves are supported in the atmosphere, resulting in more general
acoustic-gravity spectra. In the case of a magnetized atmosphere, the
magnetic fields introduces yet another restoring effect in the form of
the Lorentz force leading to a more complex system of coupled wave
phenomena collectively referred to as magneto-acoustic-gravity or MAG
waves. Apart from the gaseous envelope that surrounds a stellar
object, these waves can also occur in the radiative interiors of
cool-stars, where the criterion for the existence of such waves are
equally satisfied.

In the solar atmosphere, IGWs have been observed and are thought to
contribute to the overall energy budget
\citep{2008ApJ...681L.125S}. 
IGWs are also searched for in the deep radiative interior of the Sun
where they form standing waves called $g$-modes
(see 
\citealt{2010A&ARv..18..197A}, 
for a review, or
\citealt{2017A&A...604A..40F}, 
for more recent work). Particularly, the latter have been 
subject of active research due to the distinctive role they play in 
these stable regions. They are invoked to explain the rigid-body 
rotation of the solar interior as a result of angular momentum 
redistribution by them
\citep{1993A&A...279..431S,
1997ApJ...475L.143K, 
1997A&A...322..320Z, 
2002ApJ...574L.175T,
2008MNRAS.387..616R}.
In addition to that, their role in the  mixing of chemical species and
the Lithium depletion are still being debated
\citep{2017ApJ...848L...1R}.

The effect of magnetic fields on the propagation of IGW has also 
received attention
\citep{1998ApJ...498L.169B}.
The existence of a sufficiently strong magnetic field (10$^5$G) in the
interiors of stars may restrict certain wavenumbers from propagating
and can even alter the location of critical layers typically
associated with strong shear flow that affects IGWs. IGWs propagating
in the radiative interiors of stars can also interact with the
large-scale magnetic field, if the IGW frequency is approximately
equal to that of the Alfv\'{e}n frequency, resulting in reflection of the
waves and consequently affecting the angular momentum transport
\citep{2010MNRAS.401..191R,
2011MNRAS.410..946R}.
It also was shown that a critical field strength exist that can result
in the conversion of IGWs into other magneto-acoustic waves, which can
be used for asteroseismic measurements of the internal magnetic fields
in red giant stars
\citep{2015Sci...350..423F, 
2017MNRAS.466.2181L}.

In the same spirit, the solar atmosphere is equally interesting as it
not only harbors magnetic field on every possible scale but also
supports IGWs. The solar atmosphere is one case where the effect of
magnetic fields on IGWs may be of significance and observable. More
recent works have looked at propagating and standing IGWs in a
gravitationally stratified medium in the presence of a background
vertical magnetic field
\citep{2016ApJ...828...88H} 
and studies of the full MAG spectra has also been carried out
\citep{2013SoPh..283..383M,2016ApJ...822..116M}.

The generation mechanism of IGWs in the solar, and in general in the
stellar case (both in the interior as well as the atmosphere) mainly
fall under the category of convection driving mechanism
\citep{1990ApJ...363..694G}. 
This is in sharp contrast with the variety of different mechanisms
possible in the terrestrial case
\citep[e.g.][and references therein]{2003RvGeo..41.1003F},
which also includes orographic generation by winds passing over
mountain terrain. Convective overshooting at the location of a
convective/stable interface
\citep{1986ApJ...311..563H, 
1990ApJ...363..694G, 
2005MNRAS.364.1135R,
2013MNRAS.430.2363L, 
2010JFM...648..405A, 
2014A&A...565A..42A,
2015A&A...581A.112A, 
2016A&A...588A.122P,
2017A&A...605A..31P} 
is thought to be how IGWs are generated in the solar atmosphere. This
has been confirmed by reported observations of IGWs in the solar
atmosphere by several authors 
\citep{1991A&A...252..827K,
2003A&A...407..735R,
2008ApJ...681L.125S,
2008MNRAS.390L..83S,
2009ASPC..415...95S,
2011A&A...532A.111K,
2014SoPh..289.3457N}.

\citet[][hereafter Paper I]{2017ApJ...835..148V} 
looked at IGWs generated in model solar atmospheres and investigated
the effect of magnetic fields on their propagation. We used
state-of-the-art three-dimensional numerical simulations to study
atmospheric IGWs in the presence of spatially and temporally evolving
magnetic fields with waves naturally excited in them by convection.
The work was a significant extension to the linear analysis that was
carried out by
\cite{1982ApJ...263..386M,
1981ApJ...249..349M}. 
We considered two models, a non-magnetic and a magnetic model, and
studied the differences in wave spectra emerging from the two
atmospheres. Internal waves are generated in both models and overcome
the strong radiative damping in the lower photosphere to propagate
into the higher layers. The presence of magnetic fields show a strong
influence on these waves as they propagate higher up in the
atmosphere. We concluded that the internal waves in the quiet Sun
likely undergo mode coupling to slow magneto-acoustic waves as
described by
\cite{2011MNRAS.417.1162N,
2010MNRAS.402..386N} 
and are thereby mostly reflected back into the atmosphere.

The mode coupling scenario was further confirmed by the fact that the
mechanical flux showed a mixed upward and downward component in the
internal gravity wave regime for the magnetic case in the higher
layers while there was only an upward component in the non-magnetic
case. As the magnetic fields in this model was predominantly vertical,
conversion to Alfv\'{e}n waves was speculated to be highly unlikely.
Because of the lack of additional magnetic models to confirm 
this scenario, a study of this was not pursued in Paper I. We also 
speculated that the strong suppression of IGWs within magnetic 
flux-concentration observed by
\cite{2008ApJ...681L.125S} 
may be due to non-linear wave breaking as a result of vortex flows
that are ubiquitously present in these regions 
\citep{2012Natur.486..505W}.

Our analysis, presented in Paper I, shows that a considerable amount
of internal wave flux is produced in the near surface layers and that
these waves can couple with other magneto-atmospheric waves. Hence, it
is important to fully understand the transfer of energy from these
waves to other waves in the atmosphere of the Sun in order to account
for the complete energy sources related to wave motion. The generation
of internal waves and how magnetic fields influence their generation
in a realistic solar model atmosphere is of great interest. Whether
these waves carry sufficient energy to balance the overall energy
budget of the upper atmosphere and how the presence of magnetic field
affect the energy transport is still unclear. To this end, we present
here a study of IGWs in dynamic model solar atmospheres of different
magnetic flux density and look at gravity wave excitation and
propagation in them. We are interested in a quantitative estimate and
comparison of the energy flux associated with these waves in diverse
magnetic environments that are representative of different regions on
the Sun with magnetic flux densities that resemble regions of the 
	Sun ranging from the quiet Sun to plages.

The paper is structured as follows: In 
Section~\ref{s:numerical_models}, 
we present the numerical setup and describe
the properties of the different models. 
In Section~\ref{s:analysis}, we briefly describe the spectral 
	analysis of the 3D simulation which was already presented in 
	Paper I.
Section~\ref{s:results} presents the results of the spectral 
	analysis and looks at the differences in the the phase and energy 
	flux spectra from the different models.
Section~\ref{s:discussion}, we discuss similarities in 
	the generation and differences in the energy transport by the 
	waves in these models. The conclusion of the paper is given in 
	Section~\ref{s:conclusions}.

\section{Numerical Models}{\label{s:numerical_models}}

\begin{table*}[ht!] 
	\centering 
	\caption{Numerical setup and physical properties
		of the four simulated models.}\label{tab:model_summary}
	\label{tab:summary}
	\begin{tabular}{lcccc}
		\hline
		\hline
			& \textsf{\small Sun-v0}  
			& \textsf{\small Sun-v10}   
			& \textsf{\small Sun-v50}  
			& \textsf{\small Sun-v100} \\
			& \textsf{\footnotesize (SVGd3r05bn0cp1p\footnote{ Original identifier})}
			& \textsf{\footnotesize (SVGd3r12bv1cp1p)}   
			& \textsf{\footnotesize (SVGd3r09bv5cp1p)}  
			& \textsf{\footnotesize (SVGd3r13bvCcp1p)} \\		
		\hline
		Snapshot cadence 
		    & \multicolumn{4}{c}{30\,s}\\
		Duration of simulation 
		    & \multicolumn{4}{c}{4\,hrs}\\
		Computational grid 
		    & \multicolumn{4}{c}{480$\times$480$\times$120}\\
		Domain size 
		    &
		   \multicolumn{4}{c}{38.4$\times$38.4$\times$2.8\,Mm$^{3}$}\\
		Computational cell size 
		    &
		    \multicolumn{4}{c}{80$\times$80$\times$(50-20)\footnote{
				The vertical
				cell size varies from
				50\,km in the lower part of the computational domain
				down to 20\,km in the upper atmosphere.} km$^{3}$}\\
		Numerical scheme 
		    & \multicolumn{4}{c}{HLL-MHD} \\
		Reconstruction scheme 
		    & \multicolumn{4}{c}{FRweno}  \\
		Initial field, $B_{z}$ (uniform) 
		    & 0~G 
		    & 10~G 
		    & 50~G 
		    & 100~G \\
		Temperature, $T_{\rm eff}$ 
		    & 5759.2$\pm$3.7\,{\rm K} 
		    & 5760.1$\pm$3.6\,{\rm K} 
		    & 5766.8$\pm$3.4\,{\rm K} 
		    & 5775.0$\pm$3.3\,{\rm K} \\ 
		Intensity contrast, $\delta I_{\rm bol}^{\rm rms}$
		    & {15.38}$\pm$0.09\,\% 
		    & {15.27}$\pm$0.10\,\% 
		    & {14.82}$\pm$0.10\,\% 
		    & {14.32}$\pm$0.10\,\% \\ 
		\hline
	\end{tabular}
\end{table*}  

We use numerical simulations to mimic the dynamics of a small region
near the surface of the Sun from where the waves likely emanate. The
simulated domain is a thin slab that encompasses a small portion of
the convective interior as well as the radiative atmospheric layer of
the star. We carry out the full forward modeling of this near-surface
solar magnetoconvection using the {CO$^{\rm 5}$BOLD}\footnote{
CO5BOLD (nicknamed COBOLD), is the short form of 
``COnservative COde for the COmputation of COmpressible COnvection 
in a BOx of L Dimensions with l=2,3''. In this work we use l=3.} code
\citep[\textsf{\small version 002.02.2012.11.05e};][]{
2012JCoPh.231..919F}
in its box-in-a-star setup. The code solves the time-dependent
non-linear magnetohydrodynamics (MHD) equations in a Cartesian 
box with an external gravity
field and taking non-grey radiative transfer into account.

The radiative-MHD code, {CO$^{\rm 5}$BOLD}, has been extensively
used for various applications that model the  solar, stellar and
sub-stellar atmospheres. This includes, but is not limited to,
chromospheres of red giants
\citep{2017A&A...606A..26W},
AGB stars
\citep{2017A&A...600A.137F}
to magnetic field effects in white dwarfs 
\citep{2015ApJ...812...19T} 
and small-scale magnetism in main sequence stellar atmospheres 
\citep{2018A&A...614A..78S}
to brown dwarfs 
\citep{2013MmSAI..84.1070F}.  
{CO$^{\rm 5}$BOLD} has proven to be a valuable tool in studies
including the determination of Oxygen and Lithium abundance in cool
stars
\citep{2008A&A...488.1031C,
2017A&A...604A..44M}.
It has also been used for local helioseismology studies as well as to
study waves in the solar atmosphere
\citep{2007AN....328..323S,
2008ApJ...681L.125S,
2012A&A...542L..30N,
2011ApJ...730L..24K,
2016ApJ...827....7K}.
The code allows for the continuous simulation of solar- and stellar
convection for long time periods making it a convenient tool to
investigate waves that naturally occur at the interface between a
convective and stable environment.

In the first part of this work (Paper I), we discussed two 
different simulations, viz. a hydrodynamic and a magnetohydrodynamic 
run that were carried out using the {CO$^{\rm 5}$BOLD} code. We 
studied the differences in wave spectra 
emerging from the two physically disparate surface convection 
regimes, one without and the other with magnetic field. 
Apart from the differences in the non-linear
equations solved, the other notable difference between the two runs
were in the numerical solvers that were used. The hydrodynamic run was
carried out with a Roe solver and the MHD run used a 
HLL (Harten, Lax \& van Leer) solver.
It was seen that the HLL-MHD numerical solver is more diffusive
compared to the Roe solver which resulted in a significant difference
in the size of granules between the hydrodynamic and
magnetohydrodynamic simulated models. Despite these differences, the
overall wave spectra that originated from the surface convection of
both models were found to be the same.

In this work, we carry out a set of four new simulations of the solar
convection that includes a non-magnetic and three magnetic cases, as
explained later. In order to make a more precise comparison, we now
use the same solver
\citep[HLL-MHD;][]{2005ESASP.596E..65S,2013MSAIS..24..100S} 
for all runs in this work. Additionally, instead of using the PP
(Piecewise Parabolic) reconstruction scheme as in the previous study
(Paper I), we now use the more robust FRweno scheme
\citep{2013MSAIS..24...26F}, 
which is a combination of PP  and WENO (weighted essentially
non-oscillatory) reconstruction methods. The FRweno reconstruction
scheme is applied in the horizontal coordinates, but switches to the
van Leer scheme in the vertical coordinate due to the non-equidistant
nature of the computational grid in that direction.

All four models start with a three-dimensional (3D) state taken from a relaxed convection
simulation without magnetic field carried out using the 
{CO$^{\rm 5}$BOLD} code. The computational domain is 
38.4 $\times$ 38.4 $\times$ 2.8 Mm$^3$, discretized on 
480 $\times$ 480 $\times$ 120 grid cells,
extending $\sim$1.5~Mm below the mean Rosseland optical depth
$\tau_{R}=1$. The atmosphere wherein the waves propagate extends over
a height of $\sim$1.3~Mm above this level. The non-magnetic run
(\textsf{\small Sun-v0}) is computed by setting the initial magnetic
flux density to zero. The three magnetic runs, viz, 
\textsf{\small Sun-v10},
\textsf{\small Sun-v50}, and \textsf{\small Sun-v100}
are computed by embedding the initial model with a uniform vertical
field of 10 G, 50~G, and 100~G, respectively, in the entire domain.
All the four models are advanced for a timespan of 60 minutes to
adjust to the new solver and to redistribute the imposed magnetic
fields. Starting with this snapshot as the new initial model for the
four runs, we advance the models for 4 hours taking a 3D snapshot
every 30 seconds. A summary of the simulation setup is shown in
Table~\ref{tab:summary}.

The boundary conditions that we use in this new set of runs are
similar to the ones that were used in Paper I. As is normally the case
with side boundaries for studying waves, we use periodic boundary
conditions on the velocity, radiation and the magnetic field. Periodic
boundary conditions have the least interference with waves that
propagate within the computational domain. The top boundary is open
for fluid flow and outward radiation and the density is set to drop
exponentially into the ghost cells. The bottom boundary is also open
with a condition on the specific entropy of the in-flowing material as
described in Paper I. The magnetic fields are forced to be vertical at
the top and bottom boundary by setting the vertical component of the
magnetic field constant and the transverse component zero across the
boundary into the ghost cells.

An equation of state that adequately describes the solar plasma
including partial ionization effects is provided in a tabulated form.
The radiative transfer proceeds via a opacity binning method, with the
help of five opacity groups, adapted from the MARCS stellar atmosphere
package
\citep{2008A&A...486..951G}.
For the radiative transfer we use long-characteristics along a set of
8 rays. An Alfv\'{e}n speed-limiter with a maximum Alfv\'{e}n speed of 
40 km s$^{-1}$ is used to make the time-stepping within 
computationally feasible limits. The effect of this is not significant 
on the waves that we study and will be discussed later in the 
appropriate section.

\section{Analysis}{\label{s:analysis}}

We carry out a spectral analysis of the 3D simulations by Fourier 
	transforming the physical quantities in both space and time to 
	identify and seperate IGW from other types of waves present in the 
	domain. The fact that we are not dealing with a uniform medium 
	(as the quantities vary with height) is for the present study 
	ignored, treating the medium as locally homogeneous in the 
	horizontal direction. All the physical variables are decomposed 
	into their Fourier components in the horizontal directions and in 
	time for each	grid point in the $z$-direction.
	Although we use only 4 hour long duration of simulation, it gives 
	us adequate spectral resolution to identify waves. A tapered 
	cosine window is used for apodization in the temporal direction to 
	account for the abrupt start and end of selected segment. The 
	spatial direction is not apodized as we have used a periodic 
	boundary condition for the lateral sides in the simulation. The 
	Fourier synthesis is carried out using the Fast Fourier Transform 
	(FFT) algorithm and the derived Fourier components are then 
	represented on a $k_{h}-\omega$ diagram for each height level by 
	azimuthally averaging over the $k_{x}-k_{y}$ plane. A detailed 
	description of the analysis is given in Paper I for further 
	reference.

\section{Results}{\label{s:results}}

\subsection{Phase and coherence spectra}{\label{ss:phase_spectra}}

IGWs are identified and distinguished from acoustic waves by 
	their properties in the $k_{h}-\omega$ dispersion relation. In a 
	compressible stratified non-magnetic medium, the two types of 
	waves are separated in to two distinct branches with a band of 
	evanescent disturbance separating them in the diagnostic diagram. 
	A characteristic feature that differentiates the two waves is the 
	opposite nature of their phases for waves propagating in the same 
	direction. Phase difference spectra (or phase spectra) computed 
	from the cross-spectrum of the velocity measurements at two heights in 
	the atmosphere clearly reveal this difference in the 
	$k_{h}-\omega$ diagnostic diagram. In what follows, we shall 
	compare the four models with respect to the phase spectra they 
	show. We select two representative pairs of heights, with the 
	first pair in the lower atmosphere close to where these waves are 
	thought to be generated and a second pair higher up in the 
	atmosphere in order to study their propagation. The 
	velocity-velocity ($v-v$) phase spectra are determined from the
	vertical component of the velocity for these two pairs of 
	heights.

\begin{figure*}[ht!]
	\centering
	\includegraphics[width=1.\linewidth]{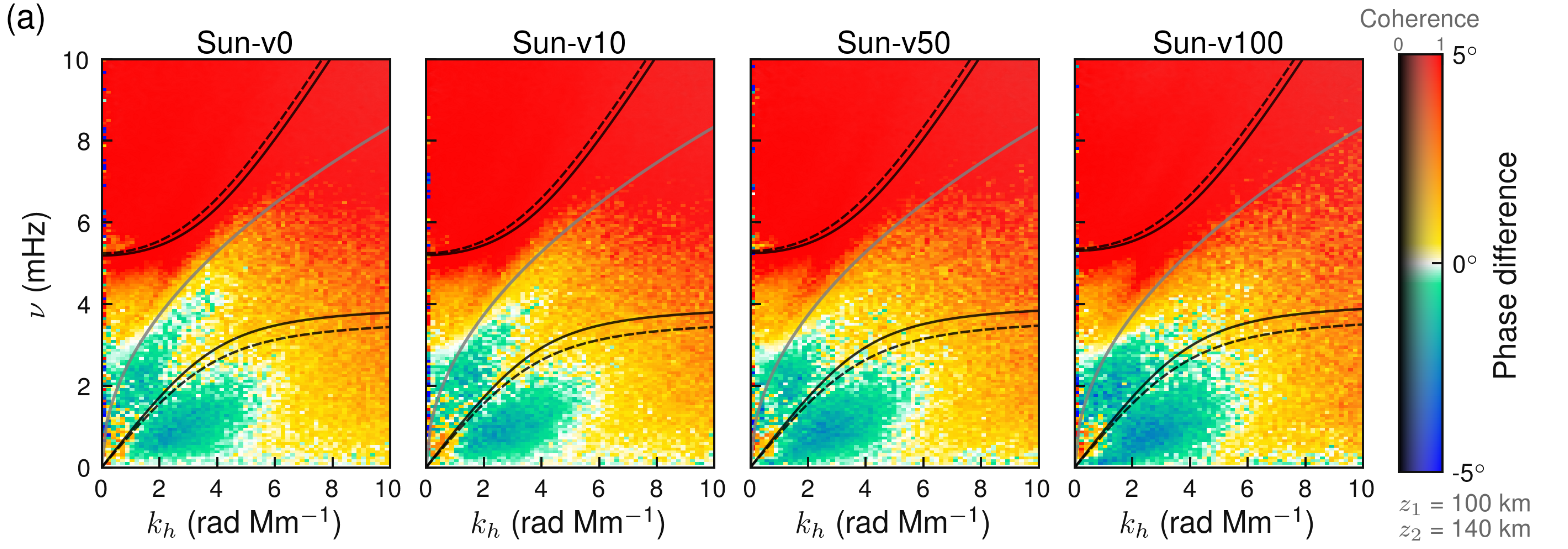}\\
	\includegraphics[width=1.\linewidth]{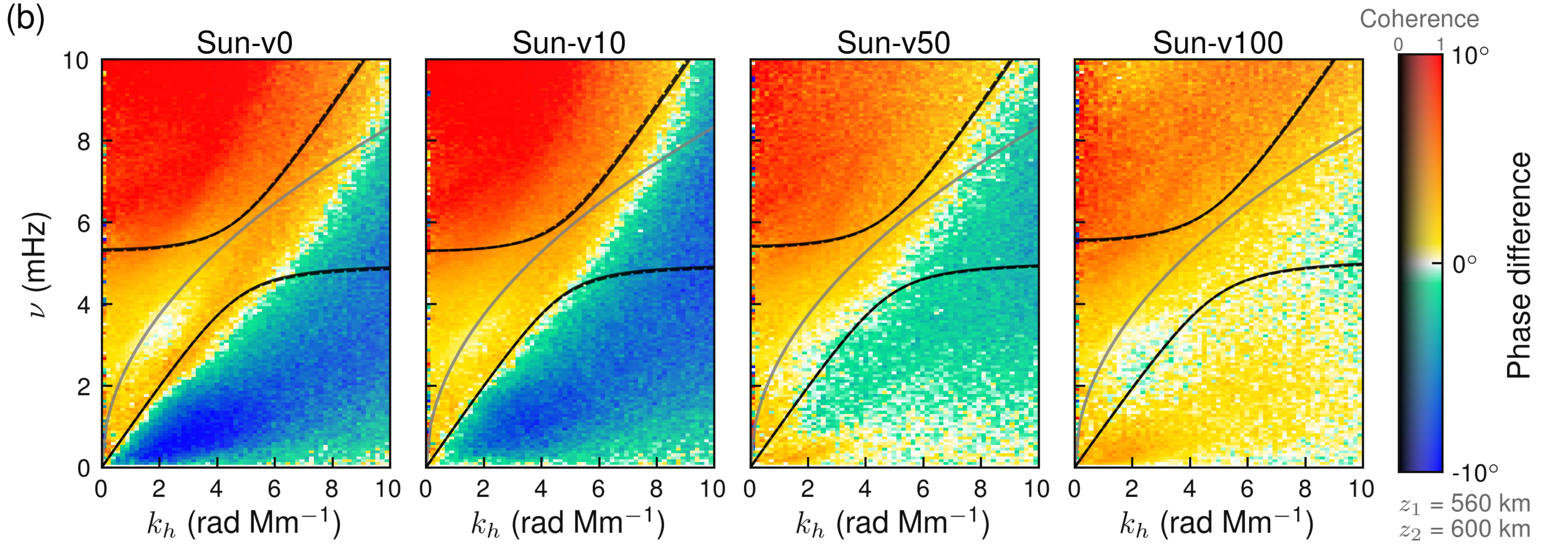}
	\caption{$v_{z}-v_{z}$ phase and coherence spectra estimated 
		between: (a) $z=100$~km and $z=140$~km and (b)
		$z=560$~km and $z=600$~km
		for the non-magnetic 
		(\textsf{\footnotesize Sun-v0}), and the three magnetic 
		models, \textsf{\footnotesize Sun-v10}, 
		\textsf{\footnotesize Sun-v50}, and 
		\textsf{\footnotesize Sun-v100}. The dashed black curves 
		represent the propagation boundaries for the lower height and 
		the solid curves those for the upper height. The gray curves 
		is the dispersion relation of the surface gravity waves. The 
		colors represent the phase difference ($\phi$) and the shading 
		shows the coherency ($\cal{K}$). IGWs propagate in the region 
		below the lower propagation boundaries.}
	\label{fig:phase_diff}
\end{figure*}

Firstly, we look at the pair of heights, $z=100$~km and 
	$z=140$~km for all the four models in 
	Figure~\ref{fig:phase_diff}(a).
	The boundary that separates the two different regimes of wave 
	propagation are shown as black curves. The dashed black curves 
	represent the propagation boundaries computed from the dispersion 
	relation for the lower height and the solid curves are those for 
	the upper height. Without the effect of magnetic fields, the IGW 
	branch of the acoustic-gravity wave spectrum occupies the lower 
	part of the diagram, below the lower propagation boundary marked 
	by black curves running from the origin to around 4mHz. In the 
	rest of the paper, we will focus on this region of the diagnostic 
	diagram. The gray curve is the dispersion relation, 
	$\omega = \sqrt{gk_{h}}$, of the surface gravity waves. It is 
	clear from Figure~\ref{fig:phase_diff}(a) that all the
	models, irrespective of magnetic fields, have the same 
	distribution in the IGWs regime shown as the greenish region below 
	the IGWs propagation boundary, showing their characteristic 
	downward phase. The only difference one can note is in the surface 
	gravity waves along their dispersion relation (marked in light 
	gray), which we will not discuss in this paper.

When we look at a slightly higher pair of heights, $z=560$~km and
	$z=600$~km as shown in Figure~\ref{fig:phase_diff}(b), we 
	notice that there are differences in the phase spectra between the 
	models, unlike what we see in the case of the lower pair of 
	heights (compare with Figure~\ref{fig:phase_diff}(a)). The
	non-magnetic model and the weak field model still show upward
	propagating (negative phase difference) IGWs, whereas the models 
	with stronger magnetic fields show downward propagating IGWs as is 
	evident from the positive phase difference (reddish region) in the 
	IGW region of the phase spectra. In summary, despite showing 
	similar phase spectra in the lower atmosphere, irrespective of the 
	average flux density, we see that the phase spectra is modified by 
	the fields in the higher layers, and shows downward propagating 
	waves in the stronger field case.

\subsection{Mechanical flux spectra}{\label{ss:mech_flux_spectra}}
\begin{figure*}[ht!]
	\centering
	\includegraphics[width=1.\linewidth]{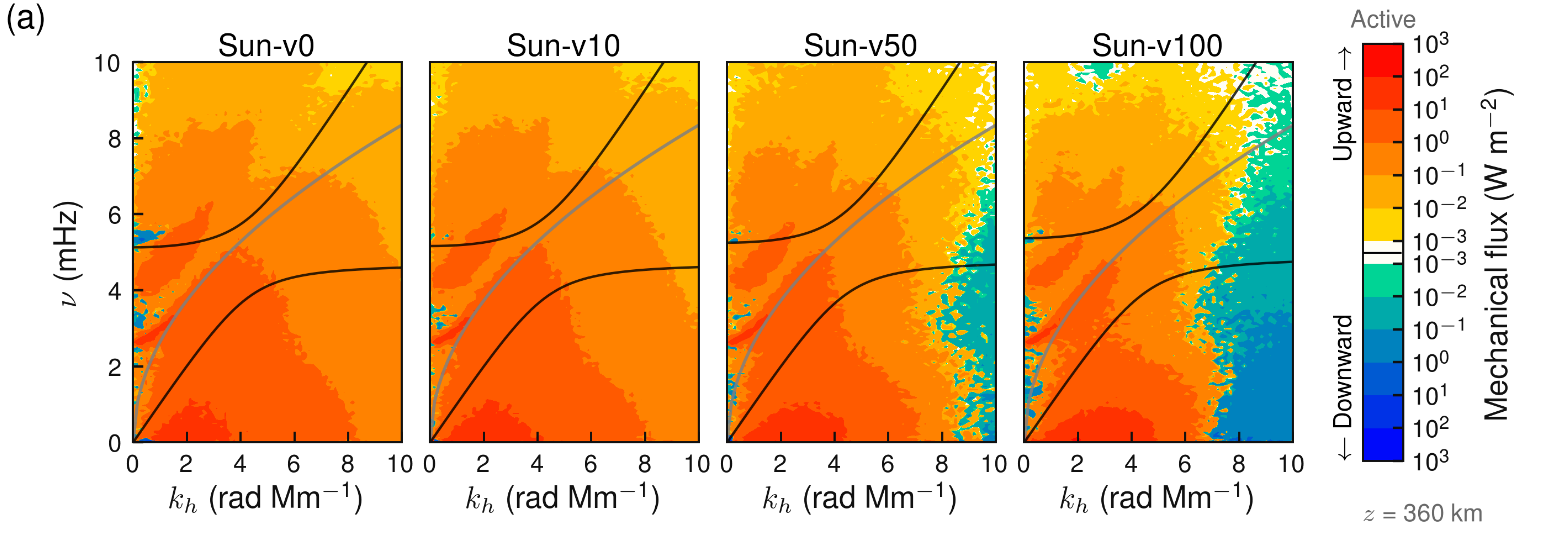}\\
	\includegraphics[width=1.\linewidth]{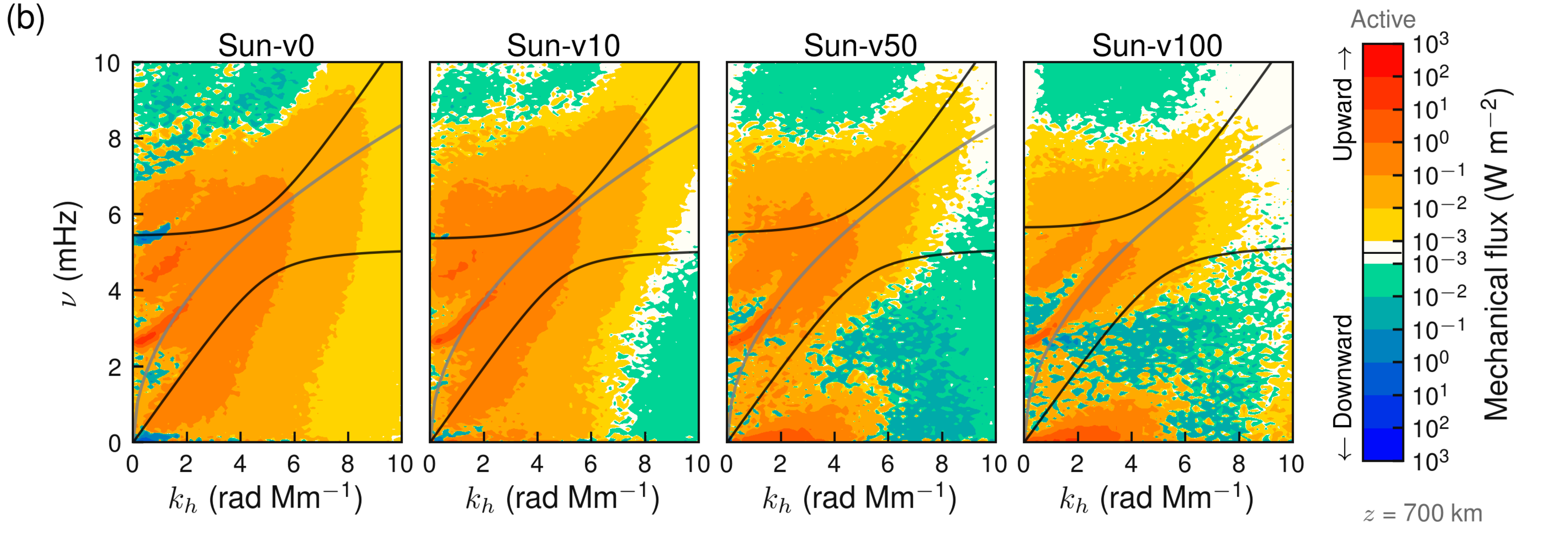}
	\caption{Vertical mechanical flux spectrum at a height of: 
		(a) $z=360$~km and (b) $z=700$~km for the non-magnetic 
		(\textsf{\footnotesize Sun-v0}) and the three magnetic models 
		(\textsf{\footnotesize Sun-v10}, 
		\textsf{\footnotesize Sun-v50}, and 
		\textsf{\footnotesize Sun-v100}).
		The black solid curves show the propagation boundaries for the
		acoustic-gravity waves. The gray curve is the dispersion 
		relation of the surface gravity waves.}
	\label{fig:mech_flux}
\end{figure*}

To understand the energy transport by waves, we will now turn our 
	attention to the energy flux spectra represented on the 
	$k_{h}-\omega$ dispersion relation diagram. The time-averaged 
	active component of the mechanical flux is given by the perturbed 
	pressure-velocity ($\Delta p$-$v$) co-spectrum (see Equation~(4) 
	of Paper I). In Figure~\ref{fig:mech_flux}, we show the vertical flux of the 
	active mechanical energy density for the four models at two 
	representative heights. We have chosen $z=360$~km and $z=700$~km, 
	in order to reveal the effects of the
	strongly varying average plasma-$\beta$ (the ratio of gas 
		pressure to magnetic pressure) with the magnetic flux density at the upper of the two heights.
	The leftmost plot in both figures each show
	the energy flux spectrum for the non-magnetic model 
	(\textsf{\small Sun-v0}), the remaining plots are for the magnetic 
	models \textsf{\small Sun-v10}, \textsf{\small Sun-v50}, and 
	\textsf{\small Sun-v100}, from left to right. 
	Examining the region below the lower propagation boundary of the 
	$k_{h}-\omega$ diagram in the lower atmosphere ($z=360$~km,
	Figure~\ref{fig:mech_flux}(a)), we see that all the fluxes 
	are positive (yellow-red colorscale). This clearly shows that the 
	energy transport is predominantly in the upward direction in the 
	order of 1-10 Wm$^{-2}$ in the given frequency-wavelength range. 
	Comparing this to Figure~\ref{fig:phase_diff}(a), we 
	clearly see here the signature of IGWs with their vertical 
	component of the phases being opposite to the propagation 
	direction. This is in contrast with the acoustic waves, occupying 
	the region above the upper propagation boundary, which show the 
	same direction for the phase and energy transport.

However, as we go higher up in the atmosphere, at $z=700$~km 
	(shown in Figure~\ref{fig:mech_flux}(b)), we notice that for 
	stronger fields, the energy flux is downwardly directed (green 
	region) in regions where the excited IGWs 
	(see Figure~\ref{fig:phase_diff}(a)) are located. The height 
	of 700~km corresponds to a low plasma-$\beta$ region for the case of strong 
	magnetic fields (for $B_{z}$ typically greater than 50~G) and 
	high-$\beta$ region for the case of weak magnetic fields, which is 
	an important difference that will be examined later in 
	Section~\ref{ss:atm_dynamics}. This shows us that as the fields 
	get stronger, there is more downward propagating IGWs in the upper 
	atmosphere transporting their energy downwards. We would like to 
	caution the reader, that for this height and the stronger field 
	cases, the propagation boundary in 
	Figure~\ref{fig:mech_flux}(b) is just shown to guide the 
	eye but otherwise has lost its physical meaning. Instead, one 
	would have to consider the complete MAG dispersion relation, which 
	is different from the acoustic-gravity dispersion relation. The 
	lower part of the $k_{h}-\omega$ dispersion relation diagram 
	correspond to the slow-MAG branch, where the flux remains upward 
	directed and it gets stronger with increasing field strength (the 
	red regions below $\nu=1$~mHz).

\subsection{Poynting flux spectra}{\label{ss:poyn_flux_spectra}}

\begin{figure*}[ht!]
	\centering
	\includegraphics[width=1.\textwidth]{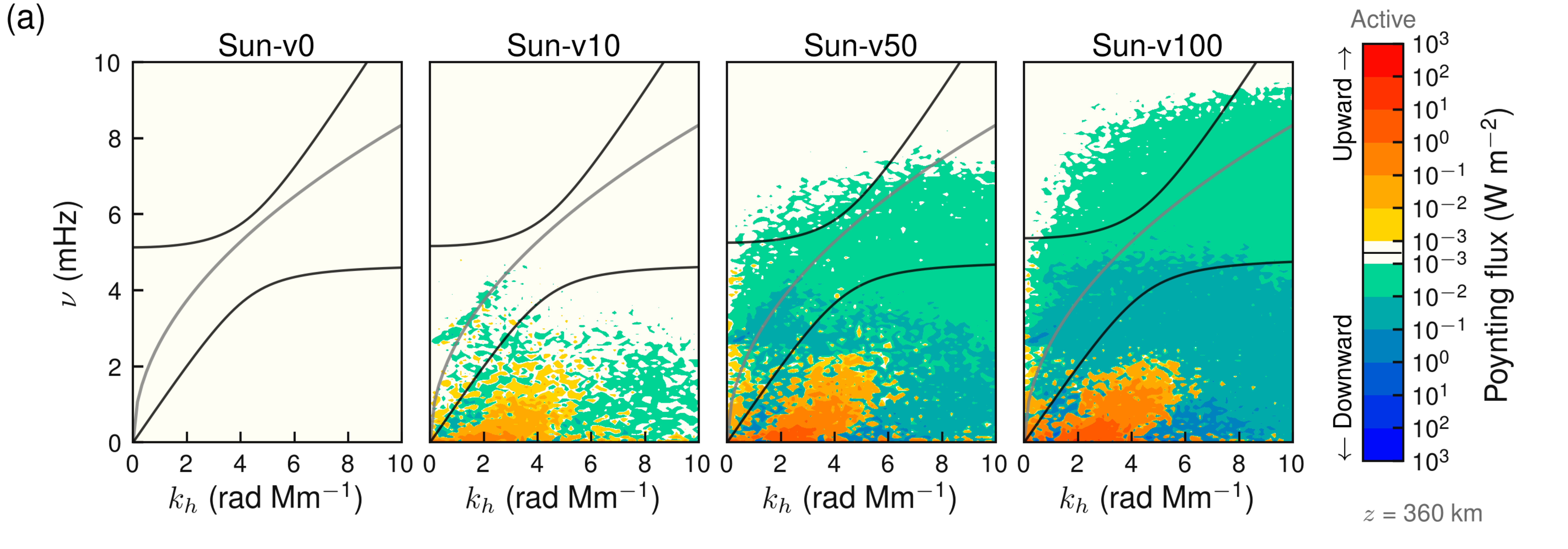}\\
	\includegraphics[width=1.\textwidth]{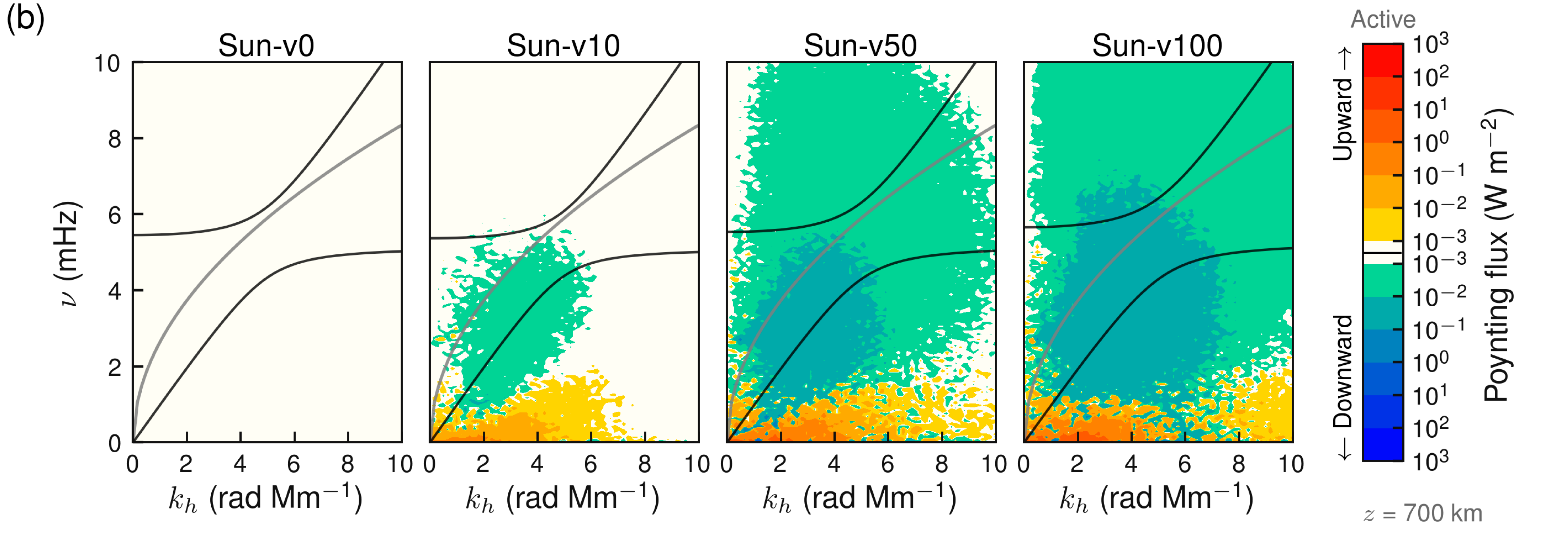}
	\caption{Vertical Poynting flux spectrum at a height of: (a) $z=360$~km and (b) $z=700$~km for
		all the models (\textsf{\footnotesize Sun-v0}, \textsf{\footnotesize Sun-v10},
		\textsf{\footnotesize Sun-v50}, and 
		\textsf{\footnotesize Sun-v100}).
		The black solid curves show the propagation boundaries for the
		acoustic-gravity waves. The gray curve is the dispersion 
		relation of the surface gravity waves.}
	\label{fig:poyn_flux}
\end{figure*}

We will now look at the active component of the Poynting
flux that can reveal more information about propagating Alfv\'{e}n 
waves in the domain.
The time-averaged flux of
electromagnetic energy is given by the real part of the complex
Poynting vector
\citep[see for e.g.][]
{1982MINTF...8.....L,
	1998clel.book.....J}.
In our case, this can be readily obtained by taking the 
	co-spectrum of the vector product of perturbed field, 
	$\bm{B^{\prime}}$, and $\bm{v \times B_{0}}$.
In Figure~\ref{fig:poyn_flux}, we show the spectra of the vertical
component of the Poynting flux from the magnetic models at two
different heights, $z=360$~km and $z=700$~km, respectively. Although,
the fluxes are a few orders of magnitude smaller than the mechanical
flux presented in Figure~\ref{fig:mech_flux}, it is interesting to look at how they 
behave as the magnetic fields become stronger and what this may tell 
us about propagating Alfv\'{e}n waves.

Looking at
Figure~\ref{fig:poyn_flux}(a), we see that for the lower
layer, which is located in the high-$\beta$ regime for all models, the
Poynting flux in the IGW/gravito-Alfv\'{e}n regime (below the lower
propagation boundary) shows a positive sign, meaning upwardly
directed. Compare this to the phase difference and the mechanical flux 
transport, both of which show that the energy is predominantly 
transported in the upward direction.

But as we go higher up in the atmosphere (see
Figure~\ref{fig:poyn_flux}(b)), we notice that these fluxes are
negative in the regions where the IGWs are excited by the convection,
which means that the Poynting flux, similar to the mechanical flux, is 
downwardly directed. This effect
gets stronger  the stronger the field (compare \textsf{\small Sun-v50}
and \textsf{\small Sun-v100}). This clearly shows us that there are
barely any upward directed Alfv\'{e}n waves. Here again, we remind 
that not the complete MAG dispersion relation is considered.

\section{Discussion}{\label{s:discussion}}
We saw that the phase spectra of the IGWs at the lower height 
remain the same irrespective of the average magnetic flux density of 
the model atmosphere. However, there are differences in the phase 
spectra as well as in both the mechanical and Poynting flux as we go 
higher up in the atmosphere. In order to explain this behavior with 
height we will explore some of the properties of the atmosphere in 
terms of the wave generation and propagation in the presence of 
magnetic field. Furthermore, we will also discuss about what this 
means for the energy transport by waves.

\subsection{Atmospheric properties}{\label{ss:atm_dynamics}}
\begin{figure*}[ht!]
	\centering
	\includegraphics[width=1.\linewidth]{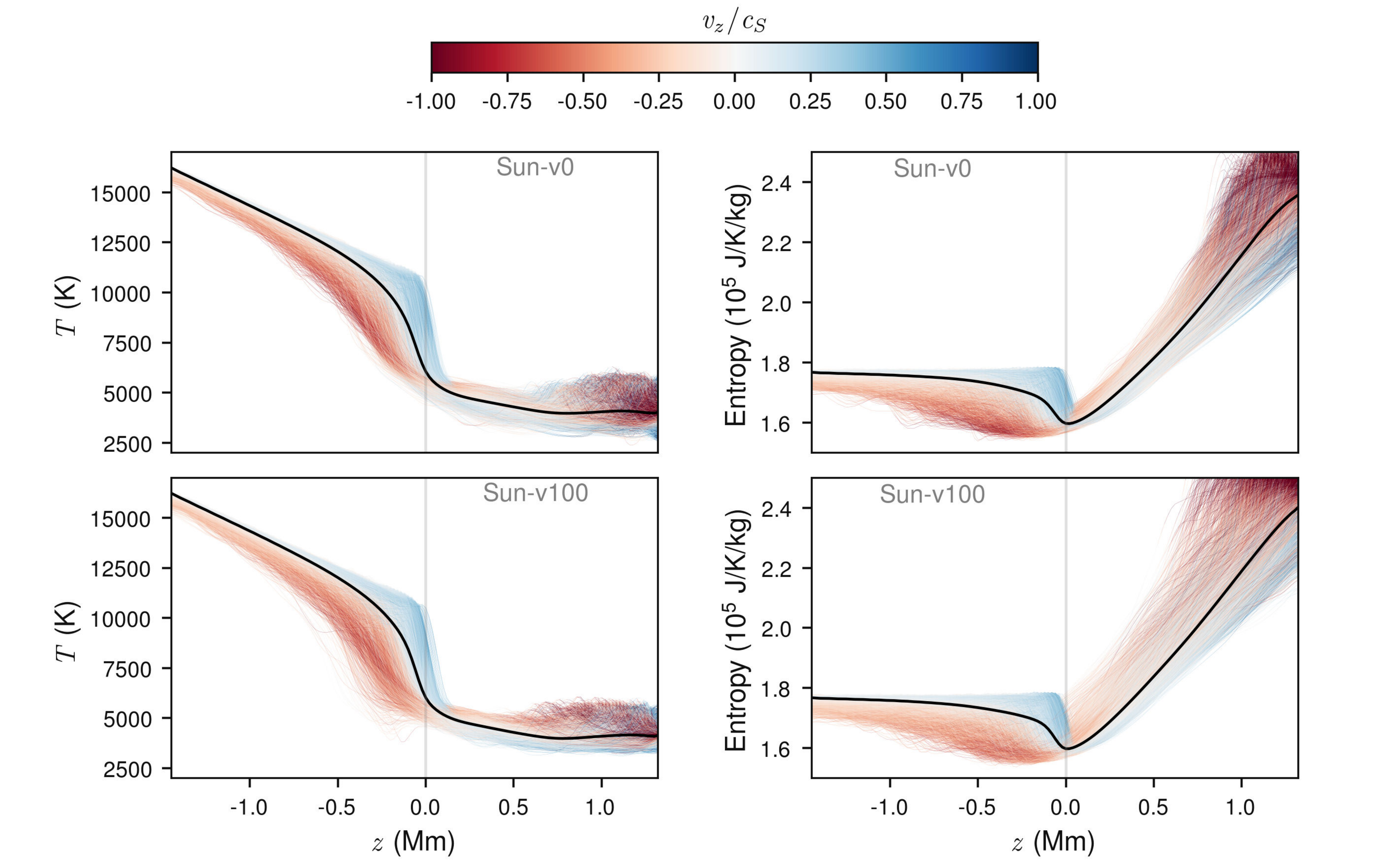}
	\caption{Temperature and specific entropy as a function of 
			height for the non-magnetic 
			(\textsf{\footnotesize Sun-v0}; top panels) and one of the 
			magnetic models (\textsf{\footnotesize Sun-v100}; bottom 
			panels). The left panels show the mean temperature (black 
			curve) run for a single snapshot taken 4 hours after the 
			start of the simulation. The colored curves show the 
			excursion of the temperature in this single snapshot. The 
			curves are color coded to show the vertical velocity 
			(scaled by local sound speed) where blue means an upward 
			velocity and red means downward velocity. The right panels 
			show the mean specific entropy (black curve) as a function 
			of height for the same snapshot as the left panels. The 
			colored curves show the excursion of the specific entropy 
			in this single snapshot. The curves are color coded to 
			show the vertical velocity (scaled by local sound speed) 
			with the colors as in left panels. The coordinate $z$ 
			increases in the outward direction.}
	\label{fig:tempvelocity}
\end{figure*}

We will first look at the physical structure and dynamics
of the magnetoconvection occurring in the different models that we
have computed. As noted in the Section~\ref{s:numerical_models}, the 
model atmospheres differ in the initial magnetic flux density that was 
added to an otherwise non-magnetic medium. We build all the models 
from the same thermally relaxed initial state, as a result of which 
the different models show very similar thermal properties. Significant 
differences are evident only when comparing the properties of the 
atmosphere where magnetic effects dominate. The effect of magnetic 
field is seen in the propagation properties of the waves only in the 
higher layers, while the generation of the waves in the near-surface 
region remain the same across the different models as evident from the 
phase and energy flux spectra presented in the previous section.

Let us first look at the thermal profile of the models that we
study here. Figure~\ref{fig:tempvelocity} shows the mean
temperature (left panels) and specific entropy (right panels) as a 
function of height (black curve) for a single snapshot taken 4 hours 
after the start of the simulation for the non-magnetic (top panels) 
and a magnetic model with the largest average magnetic flux density 
(bottom panels). In the background we show the excursion of the 
corresponding temperature and specific entropy observed at the same 
time instance over the whole domain. The curves are color-coded to 
show the vertical velocity (normalized by the local sound speed) in 
the voxels along the $z$-direction, where blue means an upward 
velocity and red means downward velocity. For clarity, the position of 
$z=0$~km height level is shown with a gray vertical line to mark the 
separation of convective and the radiative regions.

Just below the optical surface, i.e.\,below $z=0$~km, the ascension of
hot buoyant convective parcels are clearly visible as blue shade above
the mean temperature profile. Regions with lower temperatures than the
mean, appearing red, are the cool denser parcels flowing down in the
intergranular lanes. Above the surface, from $z=0$ to about 
100~km, we see that the hot voxels have overshot slightly into the 
stable layers, beyond which they radiatively cool down and are 
dominated by downward parcels as evident from the red shade. This 
overshooting hot material, that we see in all the models, are thought 
to result in the excitation of the IGWs. Further up, we see a mixture 
of upward and downward flowing plasma.

The sudden drop in temperature near the surface due
to the switch from a convective interior to a radiative atmosphere is
clearly evident in all the models. In the right panels of 
Figure~\ref{fig:tempvelocity}, we can see that the entropy rich (above 
the mean value) upwellings of the convective parcels in blue shade 
below the $z=0$~km level. Once these parcels enter the stably 
stratified radiation zone, they radiate away their heat and thereby 
loose entropy and fall back. This can be seen in the entropy-deficient 
parcels moving downward (red shade) in the convection zone. Just above 
the surface, the down falling matter is entropy rich as a result of 
heating from below. In the magnetic models, however, there is 
additional, presumably magnetic heating that results in slightly 
higher entropy of the down falling material relative to the 
non-magnetic model. The steady increase in entropy as a function of 
height in the atmosphere is visible in all models as a consequence of 
radiative absorption.
It is clear from Figure~\ref{fig:tempvelocity} that the launching 
regions of IGWs are the near-surface overshooting, which generates 
predominantly upward propagating waves above this layer. However, 
these regions are the places where we also see kG fields accumulated 
in intergranular lanes. The question is how do they influence the wave 
generation.

\subsection{Wave generation and propagation}{\label{ss:wave_dynamics}}
\begin{figure*}[ht!]
	\centering
	\includegraphics[width=1.\linewidth]{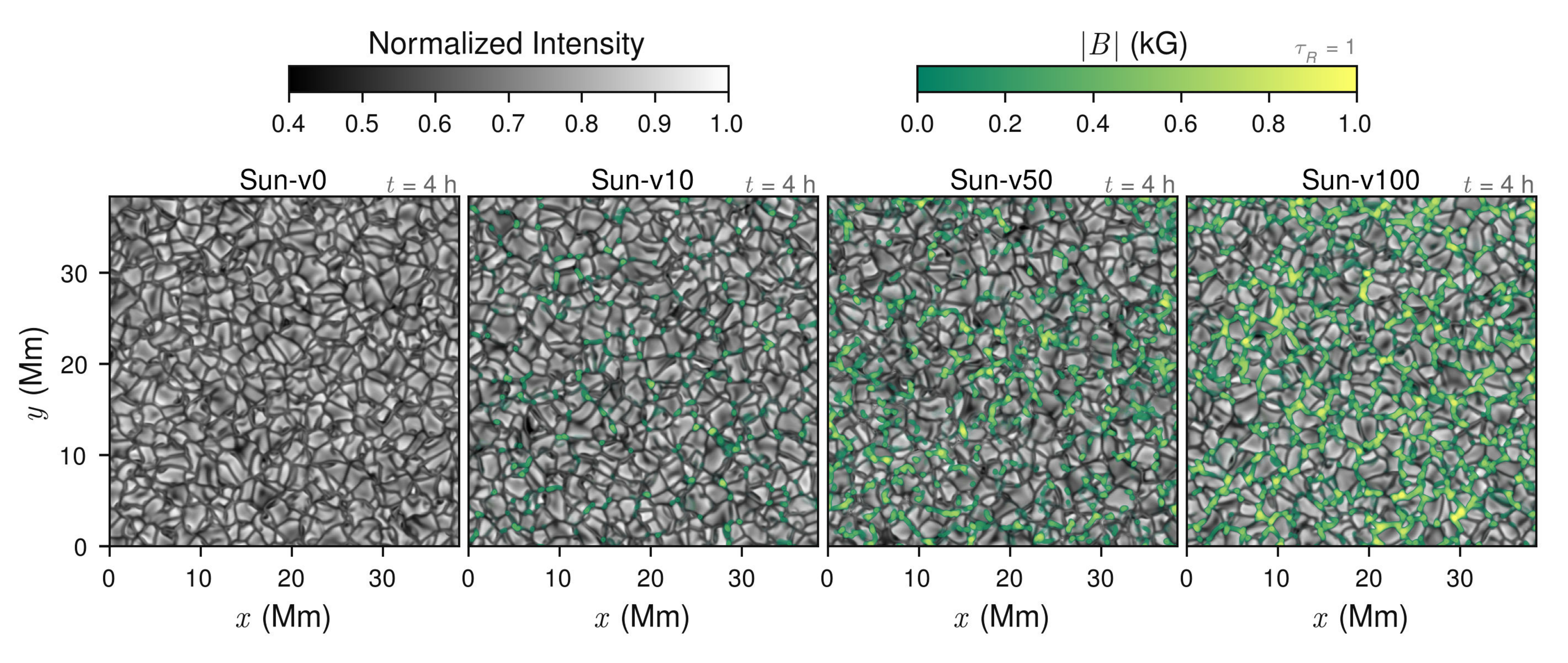}
	\caption{Emergent bolometric intensity (in grayscale) and 
			absolute magnetic field strength at $\tau_{R}=1$ (colored 
			with $\alpha$-blending) from the four models each taken 4 
			hours after the start of the simulation. An animation of 
			this figure is available online as ancillary video file.}
	\label{fig:bolometric_magnetic}
\end{figure*}

\begin{figure*}[ht!]
	\centering
	\includegraphics[width=1.\linewidth]{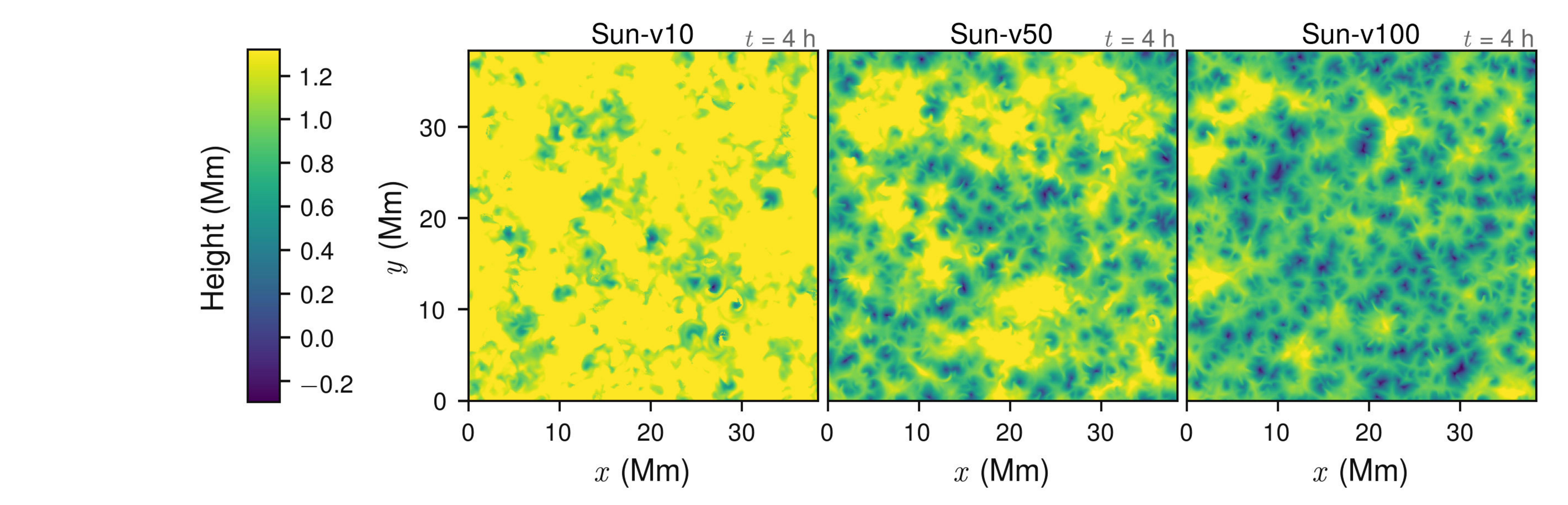}
	\caption{Height of the plasma-$\beta$=1 layer in the three 
			magnetic models from the same snapshots corresponding to 
			Figure~\ref{fig:bolometric_magnetic}.}
	\label{fig:beta_height}
\end{figure*}

Figure~\ref{fig:bolometric_magnetic} shows the emergent bolometric
intensity from the four models taken 4 hours after the start of the
simulation. Also shown is the absolute magnetic flux density in color
scale for the magnetic models. 
The accompanying animation shows the time evolution of the 
granular pattern and the magnetic map over the entire duration of the 
simulation.
One notices that the weakest magnetic model, 
\textsf{\small Sun-v10}, has very few concentrations of kG 
fluxes mostly in the vertices where several granules meet, where most 
of the fluxes accumulate due to strong convergent flows. Models with 
stronger initial field strengths have fluxes in the entire periphery 
of individual granules as well.
Comparing this with the phase spectra of the IGWs, which do not 
differ much among models, clearly shows that the kG fields that are 
present in the intergranular lanes do not have any influence on the 
generation of IGWs.

However, this is not the case for the upper layers where we do see an 
influence of the magnetic fields on the phase spectra and the energy 
flux spectra. The most important property of a magnetic atmosphere 
relevant to the study of waves is the plasma-$\beta$. For 
magneto-atmospheric waves, it is well known that the $\beta$=1 surface 
is a crucial layer across which mode coupling can occur. At locations 
of strong magnetic field the $\beta$ surface dips, so that waves 
propagate differently within small-scale flux-concentrations compared 
to the propagation in nearly field-free quiet regions. For IGWs, 
changes on these small length scales do not have a major effect, 
rather it is the average properties that matter. Since we consider 
different magnetic field strengths, a clear distinction occurs in the 
average plasma-$\beta$ of these models.
Figure~\ref{fig:beta_height} shows the height of the 
plasma-$\beta$=1 layer from the same snapshot for each models. This 
shows us the height above which the IGWs are likely to be strongly 
affected and how the height varies with varying average magnetic flux 
density.

\begin{figure}
	\centering
	\includegraphics[width=1\linewidth]{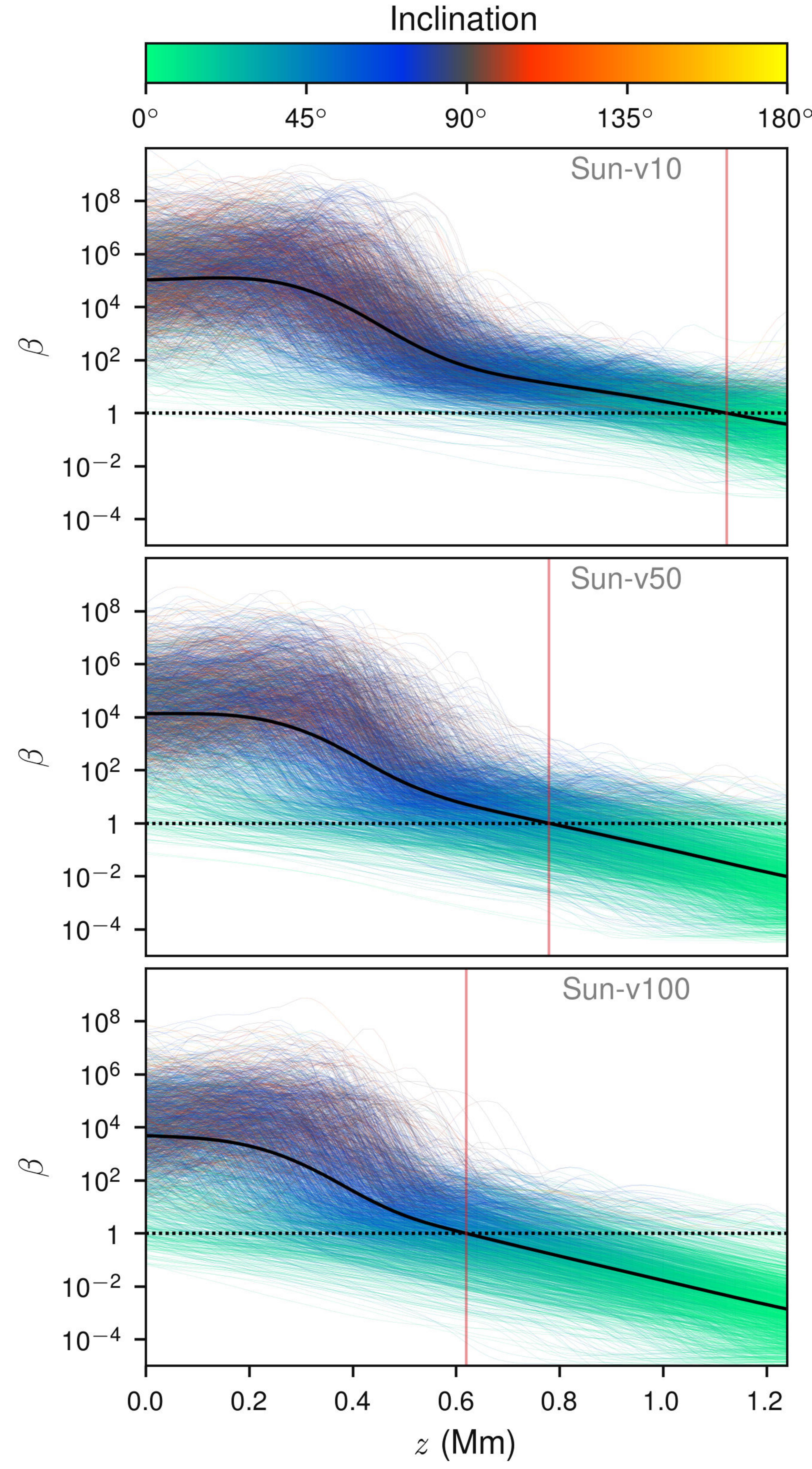}
	\caption{Average plasma-$\beta$ as a function of height (solid 
	curve) in the  three magnetic models 
	(\textsf{\footnotesize Sun-v10}, \textsf{\footnotesize Sunv50}, 
	and \textsf{\footnotesize Sun-v100}) from a single snapshot taken 
	4 hours after the start of the simulation. The excursion of 
	plasma-$\beta$ for the given snapshot is shown in the colored 
	curves, with the color representing the inclination ($\phi$) of 
	the magnetic field with respect to the vertical direction 
	according to the colorscale shown in the top. The dotted 
	horizontal line shows the $\beta=1$ value and the vertical red 
	line shows the height at which the average plasma-$\beta$ in the
	atmosphere is equal to 1, which is different for the three 
	models.}
	\label{fig:betahistogram}
\end{figure}

Figure~\ref{fig:betahistogram} shows the plasma-$\beta$ profiles as a
function of height in the three models taken from the same snapshot as
used in Figures~\ref{fig:tempvelocity}.
The average value as a function of height is marked by the black curve
and the $\beta$=1 value is marked by the dotted line to aide the
reader. The background curves show the profiles in all the domain and
are color-coded to represent the inclination of the field with respect
to the vertical direction. For the weak magnetic field case,
\textsf{\small Sun-v10}, the gas pressure dominates magnetic pressure
nearly everywhere in the atmosphere except in the very top part,
resulting in a significant portion of the atmosphere to be a
high-$\beta$ atmosphere. But as the magnetic fields get stronger, an
extended part of the atmosphere in the top layers becomes low-$\beta$.
The average plasma-$\beta$ profile moves down in the atmosphere. This
is evident from Figure~\ref{fig:betahistogram}, where we see that the
average level of $\beta=1$ is at $z=800$~km for the 
\textsf{\small Sun-v50} case and at $z=600$~km for the 
\textsf{\small Sun-v100} case, compared to $z=1100$~km for the 
\textsf{\small Sun-v10} case. These heights are indicated as red 
vertical lines to aide the reader. A consequence of this is that the 
IGWs in the low field strength case can propagate unhindered as far 
out as 1000~km, whereas for the stronger field strength case, the 
height limit is around 700~km. It is clear from
Figure~\ref{fig:betahistogram} that above the $\beta=1$ layer the
fields are mainly vertical (green shade), so that with increasing
field strength, models show an extended region of predominantly
vertical magnetic fields, which is a consequence of the top boundary 
condition.

The propagation properties of IGWs can be clearly understood in 
	terms of the phase spectra presented in 
	Section~\ref{ss:phase_spectra}. Despite the presence of kG fields 
	in the surface layers, the similar nature of the phases in the IGW 
	regime between non-magnetic and all magnetic field models, 
	confirms that the granular updrafts are the main source region of 
	these waves, as the differences between the models concerns the 
	intergranular lanes only. However, the higher layers show a 
	completely different behavior in the phase spectra, which is a 
	result of the differences in the average plasma-$\beta$ with 
	height of the three magnetic models.

\subsection{Energy Transport}{\label{ss:energy_transport}}

Let us now turn to the main objective of this paper. We are interested
in estimating the energy transported by wave-like motions,
particularly internal wave fluctuations, and in getting an insight on
how magnetic fields may influence this transport. The primary reason
why waves are considered important is because they are a carrier of
mechanical energy and momentum between the deep photosphere and the
upper solar atmosphere without the need of any material transport.
Knowing the upward flux of mechanical energy associated with the waves
as a function of height can tell us at which layer the energy gets
utilized. As we are looking at magneto-atmospheric waves, we also need
to consider the electromagnetic work done by the waves which is given
by the associated Poynting flux. The Poynting flux is then likely
converted into thermal and bulk kinetic energy by magnetic dissipation
and by the action of the Lorentz force \citep{2004ASPC..312..357S}.
In the following, we will discuss how wave energy fluxes are 
estimated from real observations and compare them to the fluxes calculated using a different approach from our simulations.

The flux estimates for various observed wave phenomena rely on the
fact that the energy flux can be determined directly from the mean
energy density and the group velocity of the wave under consideration.
Although this is generally true for any type of waves, it is often the
quantities that go into this simple expression that are prone to
uncertainties. 
Firstly, a reliable estimate of the mean energy density depends 
	on how closely the model atmosphere, from where the mass densities 
	are determined, matches the observed region. 
Secondly, the group velocity of a simple observed wave is often 
difficult to estimate, let alone for a dispersive wave for which the 
concept of group velocity breaks down. In most cases, it is not 
explicitly stated as to where the group velocity estimates come from
\citep{2012SoPh..279...43W}.

For the acoustic waves, the group velocity can be obtained using the
dispersion relation. The vertical component of the group velocity is
given as
\citep{2006ApJ...646..579F,2009A&A...508..941B},
\begin{equation}
v_{g,z} 
= \pdv{\omega}{k_{z}} = c_{s} \sqrt{1 - (\omega_{ac}/\omega)^2}
\label{eq:vgz_acoustic}
\end{equation}
which asymptotes to the local sound speed ($c_{s}$) for large
frequencies, while becoming zero for lower frequencies as one
approaches the acoustic cut-off ($\omega_{ac}$). Assuming a speed of
sound and the acoustic cut-off frequency from a model atmosphere, one
can estimate the energy flux in the desired frequency bin.
\cite{2009A&A...508..941B} 
showed that at a height of 250~km the low frequency range (5-10 mHz)
contributes more to the cumulative acoustic flux of ~3000 W m$^{-2}$
than the flux in the high-frequency (10-20 mHz) range. However,
\cite{2005Natur.435..919F,
2006ApJ...646..579F} 
report that the fluxes in the acoustic range is insufficient by a
factor of 10 to balance the radiative losses.

For the IGWs, it is quite tricky as these waves do not propagate
purely vertically. The vertical component of the group velocity
($v_{g,z}$) is given in the incompressible limit as
\citep{1981ApJ...245..286P, 
2016A&A...588A.122P},
\begin{equation}
v_{g,z} = \pdv{\omega}{k_{z}} 
= \left(\frac{\omega}{N}\right)^2\frac{\sqrt{N^2 - \omega^2}}{k_{h}},
\label{eq:vgz_igws}
\end{equation}
which can be written in terms of the vertical phase velocity
($v_{ph,z}$) as
\citep{2011A&A...532A.111K},
\begin{equation}
v_{g,z} = - v_{ph,z} \sin^{2}({\omega/N}).
\label{eq:vgz_vpz_igws}
\end{equation}

Thus, to estimate the energy flux, the main ingredient is the vertical
component of the phase velocity.
\cite{2008ApJ...681L.125S} 
determine the energy flux of the IGWs using the $v-v$ phase spectra
from 
Interferometric BIdimensional Spectrometer 
\citep[IBIS;][]{2006SoPh..236..415C} observations. The $v-v$ phase difference spectra are first
converted to phase travel time spectra and then to phase velocity
spectra under the assumption that the difference between the two line
formation heights is known. The group velocity spectra are then
determined from the phase velocity according to Equation
(\ref{eq:vgz_vpz_igws}). The energy flux spectra are finally obtained
by using the plasma density from the 
VAL model C 
\citep{1976ApJS...30....1V}
and the mean squared
velocity from Doppler shifts.
\cite{2008ApJ...681L.125S} 
found the energy flux in the IGW region to be 20800 W m$^{-2}$ at a 
height of 250 km dropping to 5000 W m$^{-2}$ at a height of 500~km in 
the atmosphere.
\cite{2011A&A...532A.111K} 
find a lower flux of  4100-8200 W m$^{-2}$ at 380 km and 700-1400 W
m$^{-2}$ at 570 km. The main reason for this discrepancy is thought to
stem from the fact that
\cite{2008ApJ...681L.125S} 
overestimate the vertical group velocity because they do not consider
the atmospheric transmission
\citep{2011A&A...532A.111K}.

With the help of numerical simulations, we can get an estimate of the
energy fluxes without needing the group velocity, since we have access
to all the physical quantities everywhere in the model atmosphere. We
can assume that the fluctuations caused by waves are weak and that
there are no strong background flows present. In such a scenario, the
non-linear MHD equations can be linearized by expanding quantities in
terms of a static background and perturbations and neglecting terms
greater than first order powers of the perturbed quantities. By
algebraic manipulation and casting the equations for perturbations in
the form of a conservation law, one can derive an energy density and a
flux of energy density associated with linear wave motion. Even for a
weak fluctuation this assumption is not valid for a long timespan as
non-linearities eventually arise due to the sole property of the
governing equations. Also, the wave energy density is conserved only
for the case of a uniform stratification
\citep{1968RSPSA.302..529B}, 
i.e for an atmosphere with constant Brunt-Väisälä frequency ($N$). 
	In the models that we study here the $N$ varies with height.
Therefore, we would like
to caution the reader that the definition of linear wave energy
density and flux does not represent the total energy density and flux
associated with a wave. To get a better estimate of the energy density
and flux, we would have to include the second order terms
\citep{1985GApFD..32..123L,
2017ApJ...837...94T}.
Having a time dependent background and spatial inhomogeneity, it is
evident that the total wave energy is not conserved. The wave can
continuously exchange energy with the background, raising concern on
casting the wave energy equation in the form of a conservation law.
Despite all these concerns, one can say that the linear energy flux
considered here gives the leading order contribution to the conserved
pseudoenergy flux
\citep{1978JFM....89..647A,1984AnRFM..16...11G}.


\begin{figure}
	\centering
	\includegraphics[width=1.\linewidth]{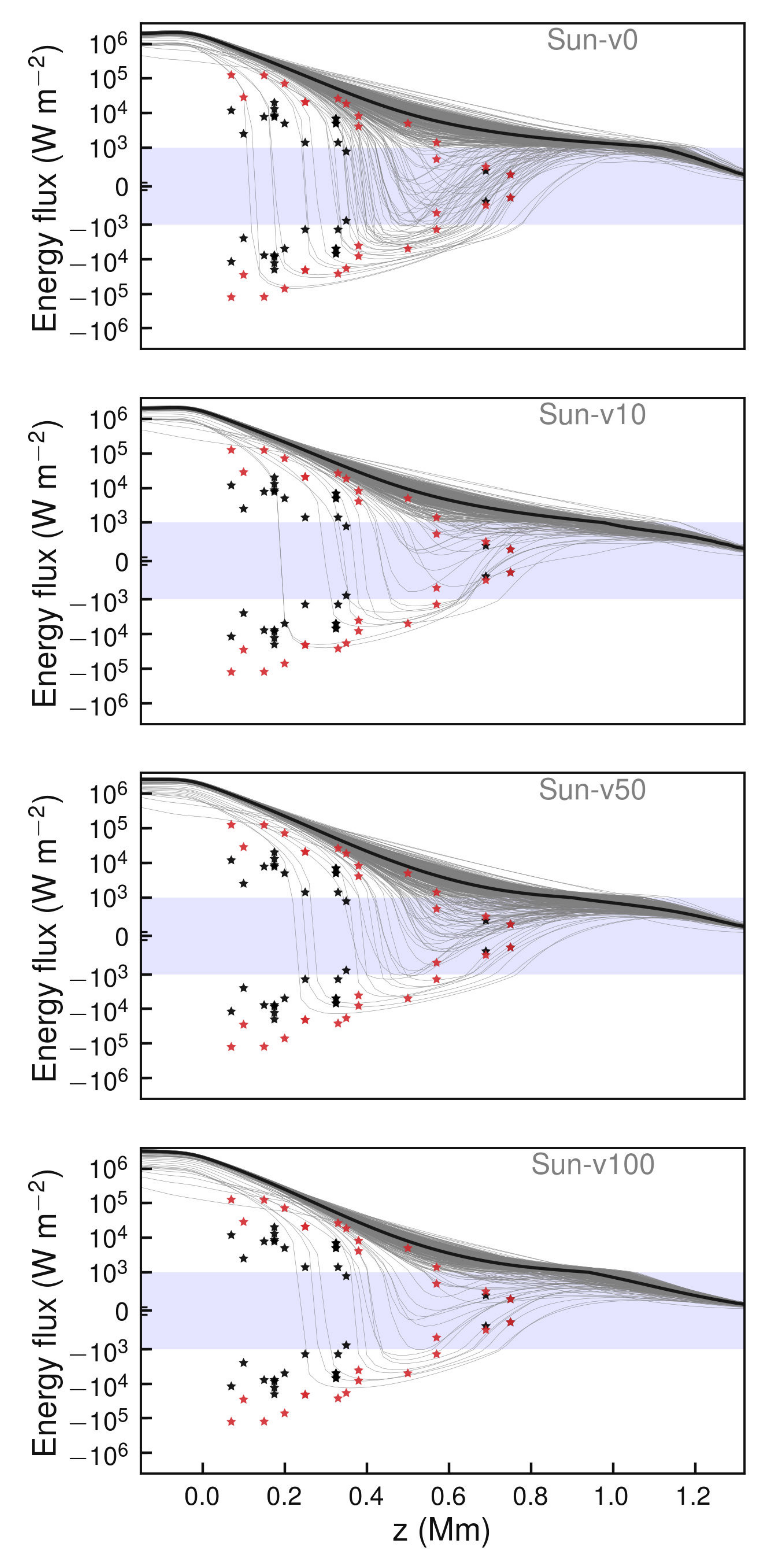}
	\caption{Horizontally averaged vertically directed mechanical 
	energy flux as a function of height for the non-magnetic
	(\textsf{\footnotesize Sun-v0}) and for the three magnetic models
	(\textsf{\footnotesize Sun-v10}, \textsf{\footnotesize Sun-v50}, 
	and \textsf{\footnotesize Sun-v100}) over the entire duration of 
	the simulation. The mean flux for individual snapshots are shown 
	in the background in gray. The asterisks denote acoustic 
	(black) and IGW (red) fluxes from observations reported in the 
	literature (see text). The blue shaded region represents linear
	scaling.}
	\label{fig:energyfluxmech}
\end{figure}

The ``linear" wave energy flux in a MHD system is made up of
mechanical energy flux and the Poynting flux,
\begin{equation}
F_{\rm{wave}} 
= \displaystyle p^{\prime} \bm{v} 
  + \frac{1}{4\pi} \qty[\bm{B^{\prime} \times (v \times B_{0})}],
\label{eq:wave_eq}
\end{equation}
where, $p^{\prime}$ is the pressure perturbation, $\bm{v}$ is the
velocity, $\bm{B_{0}}$ is the equilibrium magnetic field and
$\bm{B^{\prime}}$  is the field perturbation. The first term on
the right hand side of Equation (\ref{eq:wave_eq}) represents the rate
of mechanical work done per unit area by a volume on a adjacent volume
due to the excess pressure as a wave propagates across the volume
interface. Often quoted as the acoustic flux, in this context, we
would rather call it the mechanical flux, since it represents the
total mechanical work done by all waves, as most waves in the medium
cause pressure fluctuations as they propagate. 
This is different from the way in which the mechanical fluxes are 
estimated from observations.
The second term in Equation (\ref{eq:wave_eq}), represents the 
Poynting flux, the flux of electromagnetic energy, which is also 
carried by all the waves in the magneto-acoustic-gravity coupled 
system that we consider. There is mechanical and Poynting flux 
associated with all the waves in the medium and therefore, in the time 
domain, it is not possible to distinguish between these fluxes and map 
them to a specific wave motion. However, it is worthwhile to estimate 
the contribution to the total mechanical and the  total Poynting flux 
that arise exclusively from wave-like motions in the domain.

\begin{figure}
	\centering
	\includegraphics[width=1.\linewidth]{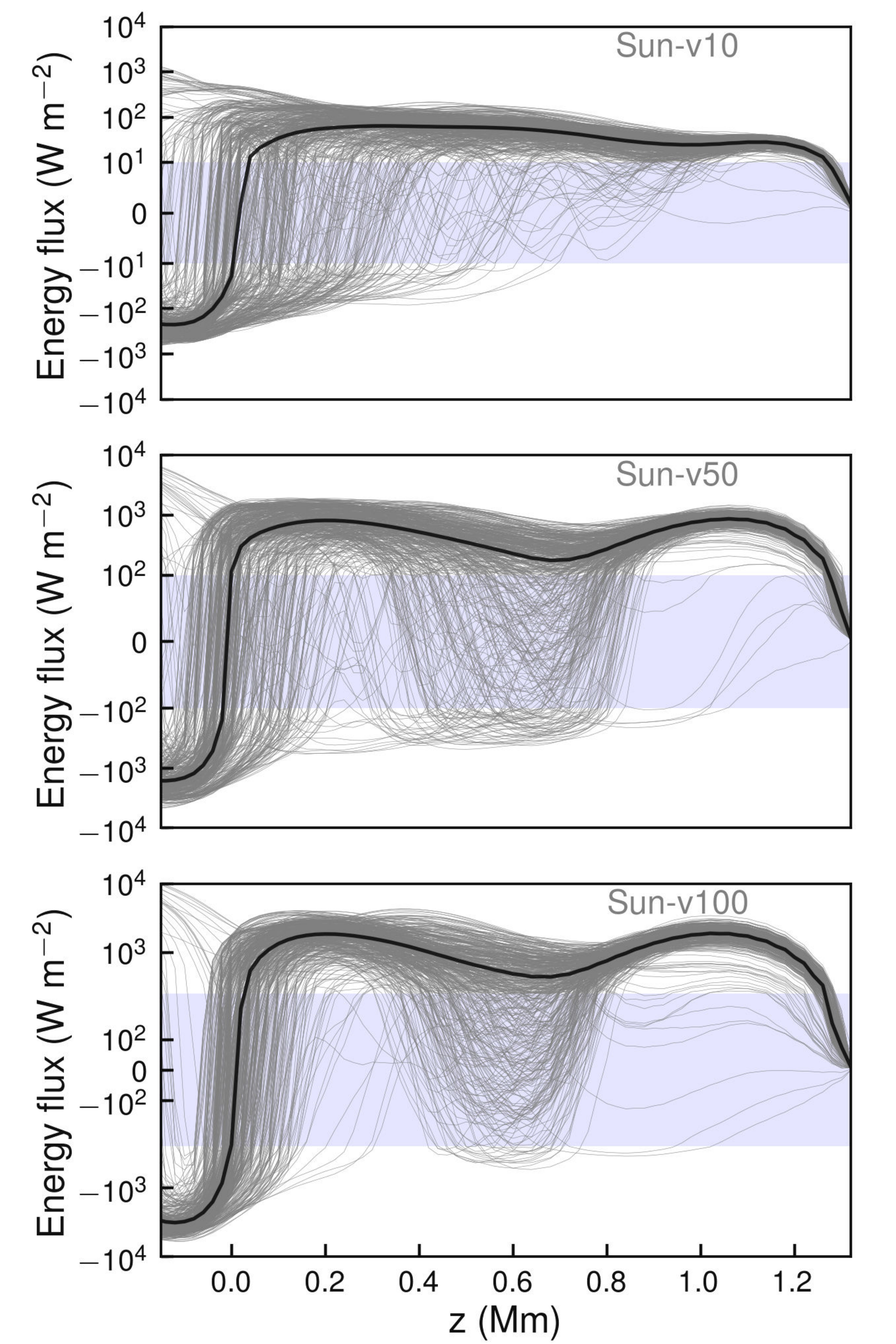}
	\caption{Horizontally averaged vertically directed Poynting 
	flux as a function of height for the three magnetic models
	(\textsf{\footnotesize Sun-v10}, \textsf{\footnotesize Sun-v50}, 
	and \textsf{\footnotesize Sun-v100}) over the entire duration of 
	the simulation. The mean flux for individual snapshots are shown 
	in gray. The blue shaded region represents linear scaling.}
	\label{fig:energyfluxpoyn}
\end{figure}

We first look at the vertical mechanical energy flux density ($F_{M}$)
as expressed by the first term in the rhs of Equation
(\ref{eq:wave_eq}). Figure~\ref{fig:energyfluxmech} shows the
horizontally averaged mean $F_{M}$ (black curve) as a function of 
height calculated from the four models. In the background we show the 
excursion of mean $F_{M}$ for individual snapshots over the duration 
of the simulation. It is clear from the height dependence of 
mechanical energy flux that we do not see much difference between the 
non-magnetic and the magnetic cases. However, we notice that there is 
a small scatter of locations with the mean flux directed downward 
(negative energy flux) in the middle part of the atmosphere around 
$z=0.6$~Mm. It is not obvious whether this scatter can be related to 
waves that are present in the box. The energy flux obtained from 
observations
\citep{2011A&A...532A.111K,2008ApJ...681L.125S,2010ApJ...723L.134B}
are marked as asterisks on all the plots. 
This includes the energy flux for the
acoustic (marked in black) as well as the IGWs (red).  As can be seen,
close to the surface the fluxes obtained from observations
($\sim10^{4}$\textendash$10^{5}$ Wm$^{-2}$) are close to an order of
magnitude less than what we see in the simulations ($10^{6}$
Wm$^{-2}$). However, this difference is less pronounced as we go
higher up in the atmosphere. Here, we also show the observed fluxes 
with the negative sign as a reference for comparing the negative 
(downward) fluxes that we obtain in simulations.

The horizontally averaged vertically directed Poynting flux 
($F_{P}$; black curve), as defined by the second term in the rhs of 
Equation (\ref{eq:wave_eq}), as a function of height is shown in
Figure~\ref{fig:energyfluxpoyn}. 
In the background we show the excursion of mean $F_{P}$ for individual
snapshots over the duration of the simulation for all the three 
magnetic models.
In all the magnetic models, the Poynting flux close to the surface 
at $z=0$~km
and 
below is directed downward as a result of the strong downdrafts in the 
intergranular lanes that carry with them magnetic flux, whereas, close 
to the surface and above, the overshooting material carries magnetic 
flux upwards resulting in a net upward Poynting flux
\citep{2008ApJ...680L..85S}.
This can be clearly seen for all three models in
Fig.~\ref{fig:energyfluxpoyn}. However, we notice that there is a
large scatter of fluxes with downward contribution. This scatter seems 
to depend on the magnetic field strength and consequently on the 
location of the mean plasma-$\beta$=1 layer. Models with a strong 
field show a downward directed Poytning flux close to the layer of 
plasma-$\beta$=1, extending over a broad height range below this 
layer. Models with weak fields, show considerably less downward 
directed Poynting flux. 
It is clear that a major contributer to the downward Poynting 
flux come from the region of IGWs as evident from the energy flux 
spectra presented in 
Section~\ref{ss:poyn_flux_spectra}.

\subsection{Limitations of the current study and future outlook}
\label{ss:limitations}

In this section, we will briefly mention some of the limitations 
of the current study and future direction. One of the major drawback 
of the present work is that the results obtained here are based on 
spectral analysis done on geometrical heights. This makes it hard to 
compare with observational data as they are mainly obtained from 
spectral maps that probe constant optical depths. A further 
exploration of this will require, performing the cross-spectral 
analysis on velocity data obtained from constant optical depths, or 
doing a spectral synthesis of different lines and performing the 
cross-spectral analysis on the estimated Dopplergrams, which will be 
explored further in a forthcoming paper.

The current simulation set-up with a rather coarse grid 
resolution of 80~km is sufficient to capture the emergent IGW spectra. 
Increasing the resolution may have an effect on the overall spectra of 
IGWs, as the granulation pattern is seen to be more structured in 
higher resolution simulation. However, due to the computationally 
expensive nature of such simulation, a study on the effect of grid 
resolution on the emergent IGW spectra was not pursued in this paper. 
In terms of boundary conditions, we have not studied the effect of how 
changing the lateral and top boundary condition will affect the 
propagation of waves. For the lateral boundary, we are restricted to 
use a periodic boundary condition as the fast Fourier transform 
algorithm assumes that the boundaries are periodic. However, imposing 
a periodic side boundary condition restricts the maximum wavelength of 
the waves to the length of the box. The top boundary condition forces 
the field to be vertical, thereby restricting us to study a magnetic 
environment that has only predominantly vertical magnetic fields. This 
however, is the case for all the models and thus makes it easy to do a 
comparative study. Further analysis with predominantly horizontal 
fields may reveal whether a coupling to low-$\beta$ gravito-Alfv\'{e}n 
waves is possible. In this case the angle of attack of these obliquely
propagating waves will be mainly parallel to the fields and therefore 
have a chance to propagate into the upper layers.

We would like to mention here the effect of the Alfv\'e{n} speed
limiter that has been used in this study. Firstly, this speed limiter
may affect waves with horizontal phase velocity greater than 40 km
s$^{-1}$ which fall mainly in the high-frequency part of the
$k_{h}-\omega$ diagram, which does not include the IGW region.
Secondly, waves with vertical phase velocity larger than 40 km
s$^{-1}$ fall close to the propagation boundary according to Eq.~1 in
Paper I and are insignificant. Therefore, we expect that the major
part of the IGW spectrum is unaffected by the use of the Alfv\'e{n}
speed limiter.
    
Finally, we would like to caution the reader that the diagnostic 
	diagram for the complete spectra of MAG waves in a uniform 
	magnetic field medium is drastically different from that of the 
	acoustic-gravity waves (considered here and in Paper I). The fact 
	that we have three distinct branches of coupled waves, viz. the 
	fast MAG, the gravito-Alfv\'{e}nic, and the slow-MAG, and that 
	their behavior in the diagnostic diagram vary depending on the 
	magnetic field inclination relative to the wave vector and 
	depending on the plasma-$\beta$, brings more variety and 
	complexity to the diagnostic diagram
	\citep[for a detailed reference see][Chapter 7, Section 7.3.3]
	{2004prma.book.....G}.
	In the low photosphere we still have a predominantly high-$\beta$
	atmosphere, which makes it easier to use the diagnostic diagram 
	for acoustic-gravity waves. As we move higher in the atmosphere, 
	we approach the surface $\beta$=1 and move into the low-$\beta$ 
	regime, where the slow- and fast-MAG branch becomes more relevant 
	and the three different branches look very different for different 
	orientation of the magnetic field relative to the propagation 
	vector. These effects are not taken care of in the present study 
	since we focus mainly on regions with high-$\beta$ where the 
	dispersion relation approximates that of the acoustic-gravity 
	waves.

More recent work by
	\cite{2017ApJ...842...37B} 
	have quoted a value of 35~G as the average magnetic flux density 
	of the quiet Sun region. Comparing it with the cases that we study 
	in this paper, this would coincide with a model intermediate to 
	\textsf{\small Sun-v10} and \textsf{\small Sun-v50}, for which the 
	average height of plasma-$\beta$=1 layer would correspond to 
	around $z=900$~km. Therefore, we argue that IGWs in the quiet Sun 
	should be easily detectable up to a height of 900~km. 
	Quasi-simultaneous measurements of 2D velocity fields at multiple 
	heights, obtained from photospheric lines, with a reasonably large 
	field-of-view (40 x 40 Mm$^2$) and an observational duration of 3 
	hours will provide a good spectral resolution in the 
	$k_{h}-\omega$ to identify the waves. Continuous observation with 
	a cadence of less than 60 seconds ($\nu_{\rm max}= 8.33$~mHz) and 
	with a moderate spatial resolution of up to 300~km 
	($k_{h,{\rm max}}=10$~rad~Mm$^{-1}$) can sufficiently capture the 
	spectral region of the emergent IGWs. We predict that the observed 
	spectra of the IGWs obtained from same spectral lines will likely 
	show differences among internetwork and the more magnetically 
	dominated plage regions. Future observations using DKIST could 
	help elucidate and validate our models.

\section{Conclusions}{\label{s:conclusions}}

Our investigation of the gravity-wave phenomena in the solar
atmosphere has shown us that these waves are naturally excited by the
overshooting convection. We study here the effect of magnetic field 
strength on the generation and propagation properties of IGWs in four 
different models of the solar atmosphere. This allows for a direct 
comparison of wave generation between magneto-atmospheres 
representative of different regions of the solar surface, like 
internetwork, network, and plage regions. We find that in the near 
surface photospheric layers the emergent IGW spectra are unaffected by 
the presence as well as by the strength of the magnetic field. IGWs 
are generated with considerable amount of wave flux independent of the 
average value of the magnetic flux density in the wave originating 
region in the quiet Sun and they propagate into higher layers. 
With increasing initial field strength, the magnetic flux tends to 
accumulate at the periphery of the granules. The main excitation 
region of IGWs are these granular upwellings into the stable region 
above.

We find that the energy fluxes obtained from observations are
	close to an order of magnitude less than what we see in the
	simulations in the lower part of the atmosphere. In terms of wave 
	energy fluxes, three conclusions can be drawn from the present 
	study. Firstly, the energy fluxes in the IGWs region are larger by 
	an order of magnitude than the fluxes in the acoustic waves in the 
	lower part of the atmosphere where plasma $\beta$ is high. 
	Secondly, the mechanical flux dominates the total wave energy 
	fluxes and is several orders of magnitude larger than the Poynting 
	flux below the plasma-$\beta$=1 level. Thirdly, the height 
	dependent horizontally averaged vertically directed Poynting flux 
	shows large scatter with a downward component below locations 
	where the average plasma-$\beta$=1. A closer look at the Poynting 
	flux in the frequency domain reveals that there is no upward 
	component for models of strong field in the IGW regime and hence 
	no mode-coupling of IGWs to Alfv\'{e}n waves across the $\beta$=1 
	surface, as a consequence of the predominantly vertical fields. In 
	the case of mechanical energy transport up to the heights where 
	the plasma $\beta$ remains greater than 1, the IGWs are quite 
	dominant and surpass the acoustic energy flux. But depending on 
	the magnetic field strength the energy in the IGWs can be 
	redirected and restricted to the near surface regions never 
	reaching higher layers. This suggests that IGWs may not be 
	abundant above the plasma-$\beta$=1 surface and therefore are 
	likely to be undetectable in observations of chromospheric layers 
	and therefore they may not directly or indirectly contribute to 
	the heating of layers above plasma-$\beta$ less than 1. In the 
	upper photosphere, however, their propagation properties depend on 
	the average magnetic field strength and therefore these waves can 
	be used as a diagnostic for the average magnetic field properties 
	of these layers.

\acknowledgements
This work was supported by the German \emph{Deut\-sche
For\-schungs\-ge\-mein\-schaft, DFG\/} grant RO 3010/3-1. 
We thank the anonymous referee for his/her detailed comments 
and suggestions which significantly improved the presentation of the 
paper.
G.V.
acknowledges the helpful discussions with Nazaret Bello Gonz\'{a}lez
and the CO$^{\rm 5}$BOLD community. The authors are grateful to the
developers of NumPy, SciPy, Matplotlib, Cython, and Astropy python
projects for providing the tools for carrying out this work. This
research has made use of NASA’s Astrophysics Data System.

\software{CO$^{\rm 5}$BOLD \citep{2012JCoPh.231..919F}, 
	NumPy \citep{walt2011numpy},
	SciPy \citep{jones2001scipy},
	Matplotlib \citep{2007CSE.....9...90H},
	Astropy \citep{2013A&A...558A..33A},
	Cython \citep{behnel2010cython}
}

\bibliographystyle{aasjournal2}
\bibliography{}

\end{document}